\documentclass[12pt,a4paper]{article}
\usepackage{amssymb}
\title{Commensurate Harmonic Oscillators: \\ Classical Symmetries}
\author{Jean-Pierre Amiet$^*$ and Stefan Weigert$^{*,**}$ \\ \\
Institut de Physique, Universit\'e de Neuch\^atel$^{*}$\\
Rue A.-L. Breguet 1, CH-2000 Neuch\^atel, Switzerland\\
and\\
Department of Mathematics, University of Hull$^{**}$\\
Cottingham Road, UK-Hull HU6 7RX, United Kingdom}
\date{November 2001}
\newcommand\be{\begin{equation}}
\newcommand\ee{\end{equation}}
\newcommand\bea{\begin{eqnarray}}
\newcommand\eea{\end{eqnarray}}
\begin{document}
\maketitle
\begin{abstract}
The symmetry properties of a classical $N$-dimensional harmonic
oscillator with rational frequency ratios are studied from a
global point of view. A {\em commensurate} oscillator possesses
the same number of globally defined constants of motion as an {\em
isotropic} oscillator. In both cases invariant phase-space
functions form the {\em algebra} $su(N)$ with respect to the
Poisson bracket. In the isotropic case, the phase-space flows
generated by the invariants can be integrated globally to a set of
finite transformations isomorphic to the group $SU(N)$. For a
commensurate oscillator, however, the {\em group} $SU(N)$ of
symmetry transformations is found to exist only on a {\em reduced}
phase space, due to unavoidable singularities of the flow in the
full phase space. It is therefore crucial to distinguish carefully
between local and global definitions of symmetry transformations
in phase space. This result solves the longstanding problem of
which symmetry to associate with a commensurate harmonic
oscillator.
\end{abstract}

\section{Introduction}
Harmonic oscillators are ubiquitous in physics. To lowest order,
motion close to a stable equilibrium of a classical system is
often described by a Hamiltonian of the form
\be \label{genosc}
H(q,p)
  = \sum_{n=1}^N \frac{\omega_n}{2}
           \left( p_n^2+q_n^2 \right) \, ,
           \quad \omega_n \in {\sf I \!  R} \, .
\ee
Here the (appropriately rescaled) canonical coordinates and
momenta have Poisson brackets $\{q_n , p_{n'} \} = \delta_{nn'}$,
$n,n' = 1 \ldots N$. If the frequencies $\omega_n$ are all equal,
\be \label{isotrop}
\omega_n = \omega \, , \quad n=1 \ldots N \, ,
\ee
the Hamiltonian (\ref{genosc}) describes an {\em isotropic}
$N$-dimensional oscillator. This system is invariant under a set
of transformations isomorphic to the group $SU(N)$: on the one
hand, the quadratic form (\ref{genosc}) in $2N$ variables is
obviously invariant under proper rotations $SO(2N)$---on the other
hand, canonical transformations need to be symplectic, hence they
are elements of $Sp(N)$. However, any transformation in ${\sf I \!
R}^{2N}$ which is both (special) orthogonal and symplectic, must
be (special) unitary \cite{perelomov90}: $SU(N) = SO(2N) \cap
Sp(N)$. The group $SU(N)$ is represented by $(N^2-1)$ phase-space
functions which, as constants of motion, generate symmetry
transformations of the Hamiltonian. In fact, the isotropic
oscillator is ``maximally superintegrable" since it possesses the
maximal number of $(2N-1)$ functionally independent constants of
motion, exceeding by far the number of $N$ globally defined
invariants required for integrability \cite{arnold}.

Suppose now that the frequency ratios $\omega_n/\omega_{n'}$ are
positive rational numbers,
\be \label{commosc}
\omega_n = \frac{\omega}{m_n} \, ,
         \qquad  m_n \in {\sf I {\!\! N_+}} \, ,
         \qquad \omega >0 \, .
\ee
This property defines a {\em commensurate} harmonic oscillator, or
$\mathbf m$-oscillator, with ${\mathbf m} = (m_1, \ldots, m_N)$.
As shown below, it also possesses $(N^2-1)$ globally defined
phase-space invariants, apart from the Hamiltonian.  Their Poisson
brackets form the Lie algebra $su(N)$, as for the isotropic
oscillator. It is known that in both systems all orbits are
closed. Nevertheless, some difference is to be expected, since all
orbits of an isotropic oscillator have the same period, while a
commensurate frequencies allow for closed orbits with different
periods. This is easily seen by exciting only individual degrees
of freedom with frequencies $\omega_n$.

In the following, the topological and group-theoretical impact of
rational frequency ratios (different from one) will be made
explicit. First, various papers dealing with commensurate
oscillators are reviewed in Section 2, which is independent of the
later developments. The technical part starts with Section 3,
where, for simplicity, the class of two-dimensional $\mathbf
m$-oscillators will be studied in detail. The generalization to $N
\geq 3$, given in Section 4, is {\em not} straightforward.
Finally, the overall picture is summarized and conclusions are
drawn. A study of {\em quantum mechanical}\ $\mathbf
m$-oscillators, including the classical limit to connect with the
present results, will be presented elsewhere \cite{amiet+02}.
\section{Symmetries of Harmonic Oscillators}
The equations of motion of $N$ harmonic oscillators can be solved
analytically for arbitrary frequency ratios. In spite of this
exceptional property many authors have wrestled with the {\em
symmetries} of such systems, the question being how their
symmetries depend on the (ir-) rationality of the frequency
ratios. Most contributions are fostered by the difficulty to
distinguish between local and global properties of phase space.
Two-dimensional oscillators with rational or irrational frequency
ratios are discussed almost exclusively. Surprising claims have
been made in the attempt to generalize properties of the isotropic
oscillator in $N$ dimensions.

{\sc Jauch} and {\sc Hill} \cite{jauch+40} address the problem of
``accidental degeneracy'' of quantum-mechanical energy
eigenvalues. The obvious invariance of the three-dimensional
harmonic oscillator (as well as the hydrogen atom) under the group
of rotations in configuration space is not sufficient to explain
the observed degeneracy of the energy levels. They conclude that
additional constants of motion must exist which account for extra
degeneracies in the quantum mechanical energy spectrum. In fact,
$(N^{2}-1)$ hermitean operators can be specified which commute
with the Hamiltonian of the isotropic harmonic oscillator in $N$
dimensions. Their commutation relations turn out to be those of
the algebra $su(N)$. Therefore, the oscillator is said to have the
$su(N)$ symmetry---which then leads to the correct degree of
degeneracies of energy levels.

{\sc Pauli} \cite{ae} and {\sc Klein} \cite{hulthen33} have
pointed out that there is a connection between degeneracies of
energy levels and the existence of further constants of motion in
the associated {\em classical} system. Therefore, the result also
should be manifest in the corresponding classical isotropic
oscillator. Upon `dequantizing' the quantum invariants, one
obtains indeed $(N^{2}-1)$ constants of motion which constitute
the $su(N)$ algebra with respect to the Poisson bracket. Hence,
the classical isotropic oscillator possesses indeed constants of
motion other than the angular momentum. Its components generate
obvious {\em geometrical} symmetry transformations while the
additional constants are said to generate {\em dynamical} symmetry
transformations. They cannot be visualized in configuration space
because they mix coordinates and momenta.

However, to exhibit a set of conserved phase-space functions which
form a particular algebra is {\em not} sufficient in order to
prove invariance of the physical system in a global sense, {\em
i.e.} in the entire phase space. {\sc Jauch} and {\sc Hill} assert
that the ``system of orbits'' of a  classical
$(m_1,m_2)$-oscillator be invariant under a group of
transformations isomorphic to the three-dimensional group of
proper rotations $SO(3)$. However, this claim cannot be justified
by local considerations only. In other words: global invariance
under a particular group of transformations does not follow from
specifying phase-space functions forming the corresponding
algebra.

{\sc Mc\-Intosh} reviews accidental degeneracy in  classical and
quantum mechanics in \cite{mcintosh59}. He notes that the phase
space of the isotropic harmonic oscillator in two dimensions
foliates into hyperspheres, being surfaces of constant energy. A
discussion of the canonical transformations generated by three
constants of the motion quadratic in the coordinates and momenta
makes follows. It becomes obvious that the group of symmetry
transformations is the special unitary group in two dimensions,
$SU(2)$---{\em not} the group of proper three-dimensional
rotations, $SO(3)$, as {\sc Jauch} and {\sc Hill} suggested.

{\sc Dulock} and {\sc McIntosh} \cite{d} devote a paper to the
two-dimensional harmonic oscillator with arbitrary frequency
ratio. Using classical variables which mimic quantum mechanical
creation and annihilation operators, they write down three
constants of motion with Poisson brackets isomorphic to the
$so(3)$ algebra relations. A Hopf mapping is performed in order to
visualize ``how the rotational symmetry of ${\cal S}^{2}$, which
is the three-dimensional rotation group, chances also to be the
symmetry group of the harmonic oscillator.'' \cite{d}. Formally,
this method can be applied to oscillators with {\em arbitrary}
frequency ratio. However, one of the transformations, which is
one-to-one in the isotropic case, becomes a multiple-valued map.
For rational frequency ratios there is a finite ambiguity, turning
to infinite multiple-valuedness if the frequencies ratios are
irrationally. In spite of this result, the authors claim that the
set of symmetry transformations for {\em all} types of oscillators
investigated is isomorphic to the group $SU(2)$---irrespective of
the multiple-valuedness. Once more, the possibility to write down
formal expressions which constitute particular algebraic relations
is taken as a proof of the existence of an associated {\em group}
of transformations.

 {\sc Maiella} and {\sc Vitale} \cite{f} react to the claim
that ``every  classical system should possess a `dynamical'
symmetry larger than the `geometrical' one'' \cite{f}.  Using
action-angle variables, they provide three constants of motion for
the two-dimensional oscillator which form the $su(2)$ algebra.
However, for irrational frequency ratio the invariants are not
single-valued---hence they consider the ``su(2) symmetry'' to be
of ``formal value'' only. It is claimed to acquire physical
relevanve only for commensurate and, {\em a fortiori}, isotropic
oscillators. At the same time, no argument is given which would
forbid the existence of the group $SU(2) $ for the irrational
oscillator.  The authors do not investigate whether, in the
commensurate case, the invariants generate indeed finite
single-valued phase-space transformations in $SU(2)$.

{\sc Maiella} \cite{g} extends this discussion to the
$N$-dimensional oscillator and emphasizes that only single-valued
constants of the motion generate actual symmetry transformations.
Initially, the group of all contact transformations for a given
dynamical system is considered. Any subgroup of transformations
which generated by single-valued constants of motion and leave the
Hamiltonian invariant, is called an ``invariance group.'' The
classical degree of degeneracy determines the number of its
generators: each linear relation between the classical frequencies
of the system with {\em rational} coefficients is accompanied by
the appearance of a single-valued constant of motion.
Subsequently, phase-space functions are given in action-angle
variables which realize the algebra $su(N)$ for an isotropic
oscillator and the algebra $su(n), \,  2 \leq n < N$, for smaller
degeneracy. However, it is again not proven explicitly that the
generators actually give rise to globally well-defined
transformations.

In the late $1960$'s, successful application of group theoretical
concepts in elementary particle physics renewed the interest in
symmetries of classical Hamiltonian systems and stimulated more
general approaches. The invariance of the three-dimensional Kepler
problem under the group of four-dimensional rotations, $SO(4)$,
was explicitly shown by {\sc Moser} \cite{ai} in $1970$ for the
first time. Already in 1965 {\sc Bacry}, {\sc Ruegg} and {\sc
Souriau} \cite{o} proved that there exists a set of global
symmetry transformations for the Kepler problem being isomorphic
to the group $SO(4)$. The transformations presented, however, do
not act on variables in phase-space. The transformations of
phase-space manifolds are parameterized by the components of
angular momentum and of the Runge-Lenz vector. Representing only
five independent constants of motion, the time $t$ at which the
particle passes the perihelion of the orbit is taken as sixth
parameter.

{\sc Dulock} and {\sc McIntosh} \cite{l} claim that the Kepler
problem has not only the symmetry $SO(4)$ but $SU(3)$. Two papers
by {\sc Bacry}, {\sc Ruegg} and {\sc Souriau} \cite{o} and by {\sc
Fradkin} \cite{p} generalize this statement: all  classical
central potential problems should possess the dynamic symmetries
$O(4)$ {\em and} $SU(3)$. This surprising statement is subject to
the same criticism as the following, even more general claim by
{\sc Mukunda} \cite{r,s}: {\em all}  classical Hamiltonian systems
with N degrees of freedom have $O(N)$ and $SU(N)$ symmetries. If
this statement were true, then there would exist just one and only
one global phase-space structure for systems with $N$ degrees of
freedom---the well-established distinction between regular and
chaotic systems would have no meaning at all.

{\sc Mukunda} argues on the basis of an a theorem by {\sc
Eisenhart} \cite{aj}. Consider, in a Hamiltonian system with $N$
degrees of freedom, $n < N$ independent functions of canonically
conjugate variables (subjected to weak conditions). They can
always be supplemented by $(2N-n)$ phase-space functions such that
$N$ pairs of canonically conjugate variables result which define a
symplectic basis of phase space. Hence, starting with the
Hamiltonian of the system under consideration one can find (i) a
variable being canonically conjugate to the Hamiltonian and (ii)
$(N-1)$ additional pairs of phase-space functions with Poisson
brackets equal to one, all commuting with the first pair and
therefore with the Hamiltonian. Consequently, this theorem is a
blueprint to construct $(2N-1)$ independent constants of motion in
any Hamiltonian system with $N$ degrees of freedom. The particular
form of the Hamiltonian does not even enter into the construction.
Next, two different sets of phase-space functions are defined in
terms of the $(2N-1)$ functions of this particular basis. Their
Poisson brackets realize the relations characteristic of the
algebras $O(N)$ and $SU(N)$, respectively. In a footnote, the
author restricts the applicability of the results: ``We concern
ourselves only with constructing realizations of Lie algebras, not
of Lie groups. Even when we talk of invariance under the $O(4)$
group, for example, we really intend invariance under the
algebra'' \cite{r}. Consequently, ``invariance under the algebra''
is a {\em local} concept only, so that  {\sc Mukunda}'s
construction has formal value only. Actually, the phase-space
functions written down by {\sc Mukunda} do not neatly map phase
space onto itself: the functions become imaginary if the range of
the canonical variables is not restricted artificially. The lesson
to be learnt is obvious: in order to establish the invariance of a
system under a {\em group} of phase-space transformations it is
not sufficient to realize specific Poisson-bracket relations with
invariants.

A related position is put forward by {\sc Stehle} and {\sc Han}
\cite{u,v}. To identify a particular algebra by constants of
motion does not guarantee the presence of  a ``higher
symmetry''---a single-valued, or at most finitely many-valued
realization of the group must exist in phase space. To show this,
they show that a system is classically degenerate if the
Hamilton-Jacobi equation of a particular system is separable in a
continuous family of coordinate systems. This property is
observable. Compare the Fourier-series representation of one
specific orbit described with respect to two different
(continuously connected) coordinate systems. For consistency, the
frequencies appearing in its Fourier decomposition must be
rationally related, which corresponds to a classical degeneracy.
It is important to note that the transformation from one
coordinate system to the other be single-valued, otherwise the
argument does not hold. Any phase-space function and,
consequently, any constant of motion generates a transformation of
phase-space onto itself; alternatively, it can be viewed as the
generator for a transition to another coordinate system such that
the Hamiltonian remains invariant. Only single-valued constants of
motion generate global single-valued transformations---infinitely
many-valued ``constants of motion'' represent formal expressions
only,  not necessarily related to the existence of classical
degeneracy. Therefore, they do {\em not} establish a higher
symmetry group of the system.

To sum up, the construction of an algebra from constants of motion
is only the first step in the proof of the existence of a
potential higher symmetry group. It needs to be supplemented by a
global investigation of the generated transformations.
\section{The two-dimensional commensurate oscillator}
This Section deals with the symmetry properties of a
two-dimensional commensurate harmonic or $(m_1,m_2)$-oscillator
described by the Hamiltonian
\be \label{twodim1}
H(q_1,q_2,p_1,p_2)
  =  \frac{\omega}{2}
      \left(  \frac{1}{m_1}(p_1^2+q_1^2) + \frac{1}{m_2}(p_2^2+q_2^2)       \right) \, ,
   \quad m_1, m_2 \in {\sf I \!  N}_+ \, ,
\ee
where the integers $m_1$ and $m_2$ have no common divisor. Two
pairs of canonical variables, $q_n,p_n \in (-\infty, \infty),
n=1,2,$ label points in phase space $\Gamma \sim {\sf I \! R}^4$,
the only non-vanishing Poisson brackets being given by
\be \label{PB1}
\{q_1, p_1\} = \{q_2, p_2\} = 1 \, .
\ee
It will be useful to introduce two other sets of canonical
variables. First, combine each pair into a complex variable
\be \label{complex2}
\alpha_n=\frac{1}{\sqrt{2}}(q_n +ip_n ) \, , \qquad n=1,2 \, ,
\ee
with non-vanishing brackets
\be \label{PB2}
\{ \bar{\alpha}_1 , {\alpha}_1\}
     = \{ \bar{\alpha}_2 , \alpha_2\}
     =  i \, ,
\ee
where $\bar{\alpha}$ denotes the complex conjugate of $\alpha$.
Second, action-angle variables $I_n \in [0, \infty)$ and
$\varphi_n \in [0,2\pi),$ $n=1,2$, are determined through modulus
and phase of $\alpha_n = \sqrt{I_n} \exp [i \varphi_n] $. Their
no-zero brackets read
\be \label{PB3}
\{ I_1, \varphi_1\} = \{I_2, \varphi_2\} = 1 \, .
\ee
These coordinates provide alternative forms of the Hamiltonian,
\be \label{hamil2}
H =\omega\left( \frac{\bar{\alpha}_1 \alpha_1 }{m_1}
       + \frac{\bar{\alpha}_2\alpha_2}{m_2}\right)
  = \omega\left( \frac{I_1}{m_1}+\frac{I_2}{m_2}\right) \, .
\ee
\subsubsection*{Constants of motion and Lie algebras}
Commensurate harmonic oscillators possess a large number of
constants of motion. The Hamiltonian itself is an invariant as $\{
H , H \} = 0$. Motion of the system with given energy $E$ is thus
restricted to a three-dimensional hyper-surface, an ellipsoid
${\cal E} (E)$ in phase space $\Gamma$. Further, the actions $I_1$
and $I_2$, having zero Poisson brackets with the Hamiltonian and
among themselves, render the $(m_1,m_2)$-oscillator integrable.
For fixed values of the actions, Arnold's theorem \cite{arnold}
states that the motion takes place on a two-dimensional torus
${\cal T} (I_1,I_2)$. In fact, the {\em entire} phase space is
foliated by tori with radii $\sqrt{I_1}$ and $\sqrt{I}_2$,
respectively. According to (\ref{hamil2}) the Hamiltonian $H$ is a
linear function of these invariants.

A third, functionally independent (complex) constant of the motion
is given by the expression
\be \label{thirdinv}
K = \alpha_2^{m_2} (\bar{\alpha}_1)^{m_1} \, .
\ee
As mentioned in \cite{jauch+40}, both its real and complex part
are invariant  which implies that the phase $\chi$ of the function
$K$,
\be \label{constangle}
\chi=m_2 \varphi_2 - m_1 \varphi_1  \in [0, 2\pi) \, ,
\ee
is a constant of the motion, too. Considered as a generator of
transformations in phase space, it connects energetically
degenerate pairs of tori. The existence of a third invariant is
expected to reduce the dimensionality of the accessible manifold.
Indeed, fixed the values of the three invariants $I_1$, $I_2$, and
$K$ (or, equivalently, $\chi$) single out a one-dimensional orbit
on the torus ${\cal T}$ if the two frequencies are rationally
related. Generic orbits, $\alpha_n(t)=\sqrt{I_n}\exp(-i\omega
t/m_n +\varphi_n(0))$, $n=1,2$, retrace themselves after a
characteristic time $t_{\mathbf m}$ $=  2\pi m_1m_2 /\omega$, with
winding numbers $m_2$ for $\alpha_1$ and $m_1$ for $\alpha_2$.
However, if the frequency ratio of the motion on the tori were
{\em not} rational, an orbit would cover the torus ${\cal T}$
densely---the function $K$ would represent a {\em formal} constant
of the motion only, without any physical impact on the motion of
the system. An important difference to the isotropic oscillator is
due to the fact that orbits of an $\mathbf m$-oscillator may have
different orbits with frequencies $\omega/(2\pi m_1)$ and
$\omega/(2\pi m_2)$, respectively. This allows to distinguish
experimentally the two cases.

The phase space of an $\mathbf m$-oscillator has a particular {\em
discrete} symmetry. Combine the variables $\alpha_n$ into a
column: now the Hamiltonian is obviously invariant under $m_1 m_2$
finite rotations, $\alpha \to R_1^{r_1} R_2^{r_2} \alpha$, $r_n =
0 \ldots m_{n}-1$, or explicitly,
\be \label{cyclicgoup}
\left(\begin{array}{c}
\alpha_1 \\
\alpha_2 \end{array} \right)
\to \left(\begin{array}{cc}
   e^{-i2\pi r_1 /m_{1}} & 0 \\
   0 & 1 \end{array} \right)
\left(\begin{array}{cc}
1 & 0 \\
0 & e^{-i2\pi r_2 /m_{2}} \end{array} \right)
\left(\begin{array}{c}
\alpha_1 \\
\alpha_2 \end{array} \right)  \, .
\ee
These transformations map the phase space $\Gamma$ to itself. They
form a cyclic group ${\cal C}_{m_1 m_2} = {\cal C}_{m_1} \times
{\cal C}_{m_2}$, the direct product of two cyclic groups with
$m_1$ and $m_2$ elements, respectively. In \cite{moshinsky+78},
${\cal C}_{m_1 m_2}$ has been called {\em ambiguity group}.

The Poisson bracket of two invariants results in a third
invariant. Therefore, the collection of all invariants is a Lie
algebra. Typically, it will contain an infinite number of
elements, all of which depend functionally on a smaller number of
invariants. By an appropriate choice of the invariants, however,
algebras with a finite number of elements can be found. The
simplest example is given by the three invariants $I_1$, $I_2$,
$K$ giving rise to the following brackets:
\be \label{invalgebra1}
\{I_1,K\}=-im_1K \, , \quad
\{I_2,K\}=im_2K \, , \quad
\{I_1,I_2\}=0 \, .
\ee
The algebra contains three independent elements---it is not
possible to find an algebra with fewer elements since the $\mathbf
m$-oscillator has three invariants. It also contains two elements
with vanishing Poisson bracket which, in a system with two degrees
of freedom, is the maximum number of `commuting' functionally
independent invariants.

There is an alternative set of four invariants $ J = ( J_0 ,
\vec{J} \, )$,
\begin{eqnarray} \label{su2gen}
J_0 & = & \frac{I_1}{2m_1} + \frac{I_2}{2m_2}
      = \frac{1}{2\omega} H \, , \\
J_1 & =& \sqrt{\frac{I_1 I_2}{m_1m_2}} \cos{\chi}  \, ,  \\
J_2 & =& \sqrt{\frac{I_1 I_2}{m_1m_2}} \sin{\chi}  \, ,  \\
J_3 & = & \frac{I_1}{2m_1} - \frac{I_2}{2m_2}      \, .
\end{eqnarray}
Only three of these invariants are functionally independent
because
\be \label{lightcone}
J_0^2 - \vec{J}^2 =  0 \, .
\ee
This constraint is conveniently rephrased by saying that the `four
vector' $ J $ is `null' or `light like.' The functions $J$ are
particularly interesting since they form the basis of a Lie
algebra isomorphic to $u(2)$,
\be \label{commsu2alg}
\{ J_0 , J_j \} = 0 \, , \qquad
\{ J_j , J_k \} =\sum_{l=1}^3\epsilon_{jkl}J_l \, ,
                                  \quad j,k=1,2,3
 \, ,
\ee
which has $su(2)$ as a subalgebra, generated by the components of
$\vec{J}$. Eqs. (\ref{commsu2alg}) has been at the origin of many
attempts to associate a group $SU(2)$ of symmetry transformations
with the two-dimensional $\mathbf m$-oscillator.
\subsubsection*{Reduced phase space and space of invariants}
Consider the complex variables
\be \label{reducedvar}
\beta_n
  =\frac{|\alpha_n|}{\sqrt{m_n}}
  \left(\frac{\alpha_n}{|\alpha_k|}\right)^{m_n}
   = \sqrt{\frac{I_n}{m_n} }\exp [ i m_n \varphi_n] \, ,
                  \qquad n=1,2 \, ,
\ee
which satisfy
\be \label{almostcanon}
\{ \bar{\beta}_n , \beta_{n^\prime} \} = i \delta_{n n^\prime} \, .
\ee
In spite of these relations, the variables $\beta_n$ do {\em not}
define pairs of canonical coordinates of ${\Gamma}$ since the map
$\alpha \rightarrow \beta$ is not a one-to-one transformation. The
variables $\beta_n$ are, however, canonical coordinates in the
{\em reduced} phase space $\Gamma_{\mathbf m}$. The reduced space
is obtained from identifying those $m_1 m_2$ points of $\Gamma$
which satisfy $\beta ( R_1^{r_1} R_2^{r_2} \alpha) = \beta (
\alpha)$, $R_n^{r_n} \in {\cal C}_{m_1 m_2}$. The definition of
the variables (\ref{almostcanon}) is motivated by the invariance
of the constants of motion in (\ref{su2gen}) under the ambiguity
group ${\cal C}_{m_1 m_2}$.

The invariants (\ref{su2gen}) take a simple form when expressed in
terms of the reduced variables,
\begin{eqnarray} \label{linearinvariants}
J_0 & = & \frac{1}{2}(\bar{\beta}_1\beta_1+\bar{\beta}_2\beta_2)
\, , \\
 J_1 & =& \frac{1}{2}(\bar{\beta}_1\beta_2 +
\bar{\beta}_2\beta_1 )
\, , \\
J_2 & =& \frac{1}{2i}(\bar{\beta}_1\beta_2 - \bar{\beta}_2\beta_1)
  \, ,   \\
J_3 & = & \frac{1}{2}(\bar{\beta}_1\beta_1-\bar{\beta}_2\beta_2)
 \, .
\end{eqnarray}
Using the two-component `Weyl spinor' $\beta = (\beta_1,\beta_2)$,
the invariants (\ref{linearinvariants}) can be written
\be \label{condenseinv}
J_{\nu} = \frac{1}{2} \bar{\beta} \cdot \sigma_{\nu} \beta \, ,
\quad \nu = 0 \ldots 3 \, ,
\ee
where $\sigma_0={\mathbf 1}_2$ and the Pauli matrices $\sigma_{k},
k=1,2,3,$ generate the algebra $su(2)$. Consequently, the
invariants, which span the space of invariants, $\Upsilon$, turn
into {\em sesquilinear} expressions on the reduced phase space
$\Gamma_{\mathbf m}$. Their structure is similar to those of the
isotropic or $(1,1)$-oscillator: formally, the reduced phase space
and the original one coincide, $\Gamma_{(1,1)} = \Gamma$. In some
sense, the non-bijective map $\alpha \to \beta$ `linearizes' the
invariants at the expense of accounting for a fraction of phase
space only. It will be shown later that the concept of the reduced
space $\Gamma_{\mathbf m}$ is natural in the present context. It
is the appropriate setting to derive global statements with
respect to symmetry transformations.
\subsubsection*{Topological aspects}
Turn now briefly to the topology of the spaces involved. Consider
the nontrivial transformations introduced so far: first, the
original phase space has been mapped to the reduced phase space,
\be \label{reduction}
\psi : \, \, \Gamma \to \Gamma_{\mathbf m} : \,
 \alpha\mapsto \beta (\alpha) \, ;
\ee
second, introducing the invariants $J$ maps the reduced variables
to the space of invariants, $\Upsilon$,
\be \label{maptoconst}
\phi:  \Gamma_{\mathbf m} \to \Upsilon : \, \, \beta \mapsto
J(\beta) \, ,
\ee
which is an upper cone in ${\sf I \! R}^4$ since $J_0=|\vec{J}|$.

The reduced phase space $\Gamma_{\mathbf m}$ has the structure of
a well-known fiber bundle. To see this, consider an orbit
$\alpha(t)$ in phase space $\Gamma$.  Its image in the reduced
space $\Gamma_{\mathbf m}$ is given by $\beta(t)=e^{-i\omega
t}\beta(0)$. The maps $\beta \to e^{i\gamma}\beta$ form a group
$U(1)$ which leaves invariant the map $\phi$, $J( e^{i\gamma}
\beta)=J(\beta)$, since the phase drops out from the sesquilinear
expressions given in Eq. (\ref{linearinvariants}). Therefore,
$\Gamma_{\mathbf m}$ is indeed a {\em fiber bundle} $( \Upsilon,
\phi, {\mathcal O})$: the invariants $\Upsilon $ form the base,
each orbit ${\mathcal O}_{\beta_0}=\{e^{i\gamma}\beta_0 \, | \,
\gamma\in [ 0 , 2\pi ) \}$ is a fiber, and the map $\phi$ is the
projection. The global structure of the bundle follows from the
fact that the restriction of $\Gamma_{\mathbf m}$ to the
sub-manifold $\Gamma_{\mathbf m}(E)$ with points $\bar{\beta}
\cdot \beta=E/\omega$ is isomorphic to the sphere ${\cal
S}^3$---as is obvious from the quadratic form (\ref{twodim1}).
Thus, the restriction of the map $\phi$ to $\Gamma_{\mathbf m}(E)$
defines the {\em Hopf fibration} of ${\cal S}^3$. To each orbit
${\cal O}_\beta$ in ${\cal S}^3$ corresponds a point
$\vec{J}(\beta)$ of the sphere of radius $J_0=E/(2\omega)$ and a
circle in the tangent space at this point.

It is interesting to look at space ${\Upsilon }$ of invariants and
the transformations among them from a general perspective. To do
so, consider the complex instead of the real Lie algebra $su(2)$
which also leaves invariant the Hamiltonian $H \sim J_0$ in
(\ref{linearinvariants}) invariant. This is the Lie algebra
$sl(2,C)$ associated with the group $SL(2,C)$, the universal
covering of the Lorentz group. The Lorentz group induced by
$SL(2,C)$ in ${\Upsilon }$ is the transitivity group of the upper
(half-) cone.

The elements of $SL(2,C)$ can be written as $u(\tau,\gamma) = \exp
[g(\tau,\gamma)]$, where $\tau$ and $\gamma$ are two real
parameters, and each $g$ is a traceless complex matrix,
\be \label{sl2c}
g(\tau,\gamma) =
       \frac{1}{2} \left( \gamma \vec{\nu} \cdot \vec{\sigma}
          + i \tau\vec{n}\cdot\vec{\sigma} \right) \, .
\ee
The matrices $u(\tau,0)$ belong to the group $SU(2)$. Thus, they
generate rotations and infinitesimal transformations can be
written in terms of a Poisson bracket:
\be \label{infsu2}
\frac{d \vec{J}(\tau)}{d\tau}
  =      \vec{n}\wedge \vec{J}(\tau)
  \equiv \{\vec{J},\vec{n}\cdot\vec{J}\} \, .
\ee

The subsets $u(0,\gamma)$ represent Lorentz boosts mapping a point
$\beta$ according to
\be \label{lorentz}
u(0,\gamma)\beta
    = \left( \cosh(\gamma/2)
           + \sinh(\gamma/2)\vec{\nu} \cdot \vec{\sigma} \right) \beta
    \equiv \beta(\gamma)
\ee
On the invariants, the transformation
\be \label{sl2cinv}
J_{\nu} = {\bar \beta} \cdot \sigma_{\nu} \beta
        \mapsto J_{\nu}(\beta)
        = {\bar \beta}(\gamma) \cdot \sigma_{\nu} \beta (\gamma) \,
        ,
\ee
is induced. Hence, the sphere ${\cal S}^3$ of radius $J_0 = H
/(2\omega)$ is mapped to a sphere of radius $J_0(\gamma)$ with
\be \label{infsl2c}
\frac{d J_0}{d\gamma}
      = \vec{\nu}\cdot\vec{J} \, , \qquad
        \frac{d \vec{J}}{d\gamma} = J_0 \vec{\nu} \, .
\ee
This is an infinitesimal Lorentz transformation which maps the
upper cone $\Upsilon \in {\sf I} \!\!{{\sf R}^4}$ to itself as is
obvious from $d(J_0(\gamma)^2 - \vec{J}(\gamma)^2)/d\gamma = 0$
and $J_0$ remaining positive. Contrary to (\ref{infsu2}), it is
not possible to express the right-hand-sides of (\ref{infsl2c}) by
means of Poisson brackets. This can be understood from a quantum
mechanical point of view. A classical theory can only manage
Boltzmann statistics whereas in quantum (field) theory, due to the
{\em anti}commutativity of Weyl spinors, it would be possible to
find a commutator to express the derivatives $d J_\nu / d\gamma$.
\section{Global invariant vector fields}
Each phase-space function generates a flow in phase space
$\Gamma$, as well as in the reduced phase space $\Gamma_{\mathbf
m}$, and in the space of invariants $\Upsilon $. The invariants
generate flows which commute with the Hamiltonian vector field. To
be more specific, consider any element $J_{\nu}, \nu = 0 \ldots 3,
$ of the Lie algebra $u(2)$. When acting on an observable $f$
through the Poisson bracket,
\be \label{vectorgen}
V_{\nu}=\{f,J_{\nu}\} \, , \quad \nu = 0 \ldots 3 \, ,
\ee
it defines a vector field $V_{\nu}$ in $\Gamma$. Its integral
lines satisfy the differential equation
\be \label{intlines}
\frac{df}{d\tau}= V_{\nu} \, .
\ee
The solution of this differential equation is a map $f(\tau)$
which will be written in the form
 \be \label{map}
f(\tau)=\mbox{Exp}[\tau J_{\nu}](f)
    = \sum_{k=0}^ \infty \{f, J_\nu \}_k \frac{\tau^k}{k!} \, ,
\ee
where
\be \label{defkbracket}
\{ g , h \}_{k+1} = \{\{ g , h \}_k,h\} \, ,
   \qquad k=1,2,\ldots \, ,   \qquad \{ g , h \}_{0} = g \, ,
\ee
with smooth phase space functions $g$ and $h$. In a simplified
notation, the solutions (\ref{map}) are written as
\be \label{sflows}
S_{\nu}[\tau] \equiv \mbox{Exp}[ \tau J_{\nu} ] \, , \quad
\nu=0\ldots 3 \, , \qquad S_{\vec{n}}[\tau] \equiv \mbox{Exp}[
\tau \vec{n} \cdot \vec{J}] \, , \qquad |\vec{n}|=1 \, ,
\ee
each unit vector $\vec{n}$ associated with a point of the unit
sphere ${\cal S}^ 2$.

The crucial question now is to investigate whether the flow
(\ref{vectorgen}) and hence the maps (\ref{sflows}) are defined
everywhere in the space under consideration. Only in this case,
the {\em algebra} formed by the closed set of Poisson brackets
among the invariants integrates to a {\em group} of symmetry
transformations. More specifically, one needs to find out whether
the invariants (\ref{su2gen}) of the ${\mathbf m}$-oscillator
generate a set of transformations isomorphic to the group $SU(2)$
(or $U(2)$). This is only possible if the associated vector fields
are well-defined everywhere in the space where they act. The
fields will be studied separately for functions $f$ from the
spaces $\Gamma, \Gamma_{\mathbf m}$, or $\Upsilon$.
\subsubsection*{Vector fields in the space of invariants}
The simplest case to look at is the orbits generated by the first
component of $J$, which is a multiple of the Hamiltonian,
$J_0=H/(2\omega)$. Not surprisingly, one has
\be \label{upsilonj0} S_0[\tau](J)=J  \, ,
\ee
that is, all components of $J$ are invariant under the action of
$J_0$.  Rotations about the 3-axis, \em i.e. \rm with an axis
passing through the poles $J_3=\pm J_0$, are generated by the
invariant $J_3=I_1/(2m_1)-I_2/(2m_2)$,
\be \label{upsilonJ3}
S_3[\tau](J)=(J_0,R_3(\tau)\vec{J}\,) \, .
\ee
Each possible orbit is generated by a linear combination of
invariants $\vec{n}\cdot\vec{J}$,
\be \label{upsilonJn}
S_{\vec{n}}[\tau](J) = (J_0,R_{\vec{n}}(\tau)\vec{J}\,) \, ,
\ee
where the matrix $R_{\vec{n}}(\tau)$ represents a rotation by an
angle $\tau$ about an axis parallel to the vector $\vec{n}$. In
other words, every point of the sphere $|\vec{J}|=J_0$ is mapped
to another point of the same sphere, the energy $E = 2\omega J_0$
being conserved.

These results are conveniently summarized by a group theoretical
statement. The set
\be \label{groupinupsilon}
{\mathcal R}_J =
         \{S_{\vec{n}}[\tau]\, | \, 0 \leq \tau<2\pi,
         \vec{n} \in {\mathcal S}^2\} \, ,
\ee
of maps acting in $\Upsilon$ is a representation of the group
$SO(3)$. In other words, there is a subset of all phase-space
functions, such that its elements transform according to the group
$SO(3)$. Mathematically, this group is the integrated form of the
adjoint representation of the algebra (\ref{commsu2alg}).
Consequently, one can attribute this group as a \em symmetry group
\rm to the reduced $(m_1,m_2)$-oscillator, for any frequency
ratio. Note, however, that this symmetry does not act on points in
phase space $\Gamma$ but on points of the space of invariants.
\subsubsection*{Vector fields in the reduced phase space}
Again, the action of the generators $J_0, J_3$, and $\vec{n} \cdot
\vec{J}$ will be studied, now with respect to the variables $\beta
= (\beta_1,\beta_2)$. It is straightforward to see that
\be \label{gammaredbeta0}
S_0[\tau](\beta) = e^{-i\tau/2}\beta \, ,
\ee
which is just the time evolution with $\tau=2\omega t$. Similarly,
the invariant $J_3$ generates a flow
\be \label{gammaredbeta3}
S_3[\tau](\beta)= \left(
\begin{array}{cc}
e^{-i\tau/2} & 0 \\
0 & e^{i\tau/2}
\end{array}\right)\beta \, .
\ee
Comparison with (\ref{almostcanon}) shows that the function
$\chi=m_2\varphi_2-m_1\varphi_1$ is left invariant. Transformation
(\ref{gammaredbeta3}) is a special case of the map
\be \label{gammaredbetan}
S_{\vec{n}}[\tau](\beta) =
 \left(
 \sigma_0 \, \cos\tau/2  - i \vec{n}
 \cdot \vec{\sigma} \, \sin \tau/2  \,  \right)
 \beta \equiv \beta(\tau) \, .
 \ee
No ambiguities arise when mapping points $\beta$ under
$S_{\vec{n}}[\tau]$, for whatever values of the parameter $\tau$
and the directions $\vec{n}$. Therefore, the set
\be \label{groupinGammared}
{\mathcal R}_{\beta} =
        \{S_{\vec{n}} [\tau]|0 \leq \tau < 4 \pi ,
         \vec{n} \in {\mathcal S}^2\} \, ,
\ee
of maps faithfully represents the group $SU_2$ in $\Gamma_{\mathbf
m}$. Consequently, an ${\mathbf m}$-oscillator admits as symmetry
not only the three-dimensional rotation group $SO(3)$ in
$\Upsilon$ but also the special unitary group $SU(2)$ in
$\Gamma_{\mathbf m}$.

In this restricted sense, and only in this one,
$(m_1,m_2)$-oscillators are seen to possess both $SO(3)$ and
$SU(2)$ as symmetry groups. This statement agrees with the fact
that the algebras $so(3)$ {\em and} $su(2)$ are isomorphic. The
next section deals with the question which groups, if any, are
represented on the original phase space $\Gamma$.

\subsubsection*{Vector fields acting in phase space}
It will be shown in this section that the vector fields associated
with the invariants $J$ are not defined globally when they act on
the variables $\alpha$ which span phase space $\Gamma$.
Consequently, it is {\em not} possible to implement the group
$SU(2)$ on phase space $\Gamma$. More explicitly, it will be shown
that the action $S_{\vec{n}}[\tau] (\alpha) = (\alpha_1(\tau),
\alpha_2(\tau))$ on $\Gamma $ is non-linear, and that it is
inevitably singular for some parameters $(\vec{n},\tau)$ and
initial points $\alpha$. Contrary to one's intuition the flows can
be defined only locally, and they cannot be extended to define a
{\em group} of symmetry transformations.

To begin with, consider the flows generated by $J_0$ and $J_3$,
respectively. The resulting orbits {\em are} well-defined for all
initial points: they are given by
\be \label{gammaJ0}
S_0[\tau](\alpha) = \left(
 \begin{array}{cc}
        e^{-i\tau/(2m_1)} & 0                \\
        0                 & e^{-i\tau/(2m_2)}
\end{array}\right)
\alpha \, ,
\ee
and by
\be \label{gammaJ3}
S_3[\tau](\alpha) =
 \left(
   \begin{array}{cc}
          e^{-i\tau/(2m_1)} & 0               \\
          0                 & e^{i\tau/(2m_2)}
   \end{array}
 \right)
\alpha \, ,
\ee
respectively. Eq. (\ref{gammaJ0}) describes the time evolution of
the point $\alpha \in \Gamma$, hence both the energy $E = 2 \omega
J_0$ and the torus ${\cal T} (I_1,I_2)$ are left invariant. Since
the values of the actions change according to $I_n(0) \to
I_n(\tau) = |\alpha_n(\tau)|^2$, the flow in (\ref{gammaJ3}) also
conserves the energy while mapping a torus ${\cal T}(I_1,I_2)$ to
another one, ${\cal T}(I_1(\tau),I_2(\tau))$.

Now consider fields which are generated by arbitrary linear
combinations of the invariants, $\vec{n}\cdot\vec{J}$. Denote
potential solutions of the differential equation
\be \label{nosolutions?!}
\frac{d\alpha}{d\tau}
  = \{\alpha ,\vec{n}\cdot\vec{J}\}
    \equiv V_{\vec{n}}(\alpha)
\ee
by $\alpha_{\vec{n}}(\tau)= S_{\vec{n}}[\tau](\alpha(0))$, with
some initial point $\alpha(0) \in \Gamma$. Explicitly, the complex
two-component field $V_{\vec{n}}$ reads
\be \label{complexfield}
V_{\vec{n}} = \frac{1}{2i}
  \left(\begin{array}{ll}
      (\vec{n}\cdot\vec{J}+im_1(\vec{n}\wedge\vec{J})_3
        + n_3J_0)/\bar{\alpha}_1 \\
      (\vec{n}\cdot\vec{J}+im_2(\vec{n}\wedge\vec{J})_3
        - n_3J_0)/\bar{\alpha}_2
\end{array}\right) \, .
\ee
It is finite but ill-defined on the hyperplanes ${\cal P}_1 =
\{\alpha \, | \, \alpha_1 = 0, \alpha_2 \neq 0 \}$ and ${\cal P}_2
= \{\alpha \, | \, \alpha_1 \neq 0, \alpha_2 = 0) \}$. There are
points which, when transported by the flow $S_{\vec{n}}[\tau]$,
hit the planes ${\cal P}_1$ or ${\cal P}_2$ for some value of
$\tau$. The associated orbits will be called {\em singular} since
they cannot be continued unambiguously across the planes. This is
due to the terms in (\ref{complexfield}) which contain $J_1$ and
$J_2$,
\be \label{weirdterms}
J_1 \pm iJ_2 = \frac{|\alpha_1||\alpha_2|}{\sqrt{m_1m_2}}
               \exp [\pm i\chi] \, ,
\ee
while all other terms are zero on ${\cal P}_1$ and ${\cal P}_2$.
Here is a toy example to illustrate the underlying problem.
Consider a one-dimensional system with variable $\alpha = \sqrt{I}
\exp [i \varphi]$, satisfying $\{ I , \varphi \} =1$. The flow
generated by $\sqrt{I}$ is ill-defined at the origin,
\be\label{toy}
\frac{d \alpha}{d \tau}
  = \{ \sqrt{I} , \alpha \}
  = \frac{i}{2} \exp [i\varphi] \, ,
\ee
as its value depends on the way the point $\alpha$ is approached.
If a trajectory were reaching the origin, it would be impossible
to continue it unambiguously beyond this point. It is important to
realize that this singularity as well as the one encountered in
the singular planes is not due to a choice of coordinates but an
intrinsic property of the flow.

To visualize the entire set of singular orbits, look at their
images in ${\Upsilon}$, that is, the orbits $S_{\vec{n}}[\tau]
(\vec{J} \, )$, $\tau \in {\sf I \! R}$. For given energy $E = 2
\omega J_0$, the points of ${\cal P}_1$ correspond to the north
pole $(0,0,J_0 = |\alpha_1|^2/2m_2 ) $ of the sphere ${\cal S}^2
(J_0)$, while those of ${\cal P}_2$ are mapped to its south pole,
$(0,0,-J_0 = - |\alpha_2|^2 / (2 m_1))$. By $\psi \circ \phi$, an
orbit $S_{\vec{n}}[\tau](\alpha)$  goes to a circle
$R_{\vec{n}}(\tau)\vec{J}(\alpha)$. {\em Singular} orbits thus
correspond to circles going through either one or both poles of
the sphere, while {\em regular} orbits hit neither of them: for
almost all flows, associated with a given vector $\vec{n}$, there
exist two ``critical'' circles passing through the north pole and
the south pole, respectively. These circles coalesce into a single
one passing through {\em both} poles if the axis of rotation is in
the equatorial plane, $\vec{n}=(n_1,n_2,0)$. They degenerate to
points located at the poles if $\vec{n}= \vec{e}_3 = (0,0,\pm 1)$.
Two conclusions can be drawn from this picture:
\begin{enumerate}
\item
for any given unit vector $\vec{n} \neq \pm \vec{e}_3$, the map
$S_{\vec{n}}[\tau]$ has at least one singular orbit in $\Gamma$;
\item
any point $\alpha \in \Gamma$ can be sent to a singular hyperplane
by a map $S_{\vec{n}}[\tau]$ with an appropriately chosen vector
$\vec{n}_{\cal P}$.
\end{enumerate}
In fact, the vectors $\vec{n}_{\cal P}$ can be chosen from {\em
two} continuous sets: they only need to be in a plane (passing
through the origin) which is perpendicular to either of the
vectors $\vec{J} (\alpha) \pm J_0 (\alpha ) \, \vec{e}_3$, or
explicitly,
\be \label{thosevectors)}
\vec{n}^{\pm}_{\cal P}
  = \frac{c_1 (\vec{J}(\alpha) \pm J_0 (\alpha) \vec{e}_3) \pm
            c_2 \vec{J}(\alpha) \wedge \vec{e}_3}
        {|c_1 (\vec{J}(\alpha) \pm J_0 (\alpha) \vec{e}_3) \pm
            c_2 \vec{J}(\alpha) \wedge \vec{e}_3 |} \, .
\ee

{\em Regular} orbits of $S_{\vec{n}}[\tau]$ are easily computed
without solving the differential equation (\ref{nosolutions?!}).
One needs to determine modulus and phase of the variables
$\alpha_n, n=1,2,$ as a function of $\tau$. It is useful to write
down the orbits in the reduced phase space and in the space of the
invariants. According to (\ref{gammaredbetan}), the reduced
variables evolve linearly,
\begin{eqnarray} \label{linearred1}
\beta_1(\tau) & = & (\cos{\tau/2} - in_3 \sin{\tau/2})
                     \beta_1 - (in_1+n_2) \sin{(\tau/2)} \beta_2
                      \, ,  \\
\beta_2(\tau) & = & (\cos{\tau/2} + in_3 \sin{\tau/2})
                      \beta_2 - (in_1-n_2) \sin{(\tau/2)} \beta_1
                      \, ,
\end{eqnarray}
while the invariants $\vec{j} = \vec{J}(\beta)$ evolve in
${\Upsilon }$ as
\be \label{linearconst}
\vec{j}(\tau)  =
    \cos{\tau} \, \vec{j} + (1-\cos\tau) \vec{n} \circ \vec{n}
          \cdot \vec{j} - \sin{\tau}\, \vec{n} \wedge \vec{j} \, .
\ee
Using $|\alpha_n|^2 = m_n |\beta_n|^2, n=1,2$, the
$\tau$-dependence of the moduli is simply
\be \label{twomoduli}
|\alpha_1(\tau)|^2 = m_1 \left( j_0 + j_3(\tau) \right) \, ,
   \qquad
|\alpha_2(\tau)|^2 = m_2 \left( j_0 - j_3(\tau) \right) \, .
\ee

For the evolution of the phases, plug Eqs. (\ref{linearred1}) into
\be \label{phasefact}
\exp [im_1 \varphi_1(\tau)]
  = \left( \frac{\alpha_n (\tau)}{|\alpha_n(\tau)|}\right)^{m_n}
  = \frac{\beta_n(\tau)}{|\beta_n(\tau)|} \, , \quad n=1,2 \, ,
\ee
giving
\be \label{phase1}
\exp [im_1\varphi_1(\tau)] =
\frac{(\cos{\frac{\tau}{2}}-in_3\sin{\frac{\tau}{2}}) \,
       \beta_1 - ((in_1+n_2)\sin{\frac{\tau}{2}}) \, \beta_2}
     {|(\cos{\frac{\tau}{2}}-in_3\sin{\frac{\tau}{2}}) \, \beta_1 -
        ((in_1+n_2) \sin {\frac{\tau}{2}}) \, \beta_2|}
    \equiv \exp[i\Phi_1(\tau)] \, ,
%
\ee
and a similar equation for $\exp [im_2\varphi_2(\tau)]$. The two
phases $\varphi_n(\tau)=\Phi_n(\tau)/m_n , n=1,2,$ must be
continuous whenever $\tau$ reaches the value $4\pi$. They will
both have a value which is a multiple of $2\pi$ when the parameter
$\tau$ takes the value $4 \pi m_1 m_2$. This result seems to
suggest that $S[\tau \vec{n}\cdot\vec{J}](\alpha)$ might be an
$m_1m_2$-fold covering of the subgroup
$S[\tau\vec{n}\cdot\vec{J}](\beta)$, $0\leq\tau < 4\pi, \vec{n}
\in {\cal S}^2$, of the special unitary group, $SU(2)$. Due to
existence of singular orbits, however, this is not possible.
Further, it is well-known that the only universal covering of
$SU(2)$ is this group itself. Nevertheless, one might describe the
situation as a {\em ramified covering} of $SU(2)$ since the maps
$S[\tau \vec{n}\cdot\vec{J}]$ combine according to a group product
law.

To visualize the obstruction of a global action of the group
$SU(2)$ differently, recall that a given map
$S[\tau\vec{n}\cdot\vec{J}]$ sends a torus ${\cal T}(I_1,I_2) \in
\Gamma$ to a torus ${\cal T} (I'_1, I'_2)$ such that
$m_1I_1+m_2I_2=m_1I'_1+m_2I'_2$ holds. For some $\vec{n}$ and
$\tau_0$ it happens that one of the actions vanishes, $I'_1$, say.
This means that the initial two-dimensional torus (${\cal S}^1
\times {\cal S}^1$) is mapped to a {\em one-dimensional torus},
{\em i.e.} a circle ${\cal S}^1$, and, therefore, one of the angle
variables has lost its meaning. Once this has happened, it is
impossible to unambiguously continue the trajectory which has hit
the singular plane, as the missing angle could take any value. The
phenomenon is similar to the passage of a spherical wave through a
focus.

It will be useful to give a name to the situation encountered
here. A system with phase space $\Gamma$ will be said to have a
{\em faint} $G$ {\em symmetry} if it admits a set of globally
defined invariants which form an algebra ${\cal A}$ while the
group $G$ associated with it cannot be realized on $\Gamma$ but
only on a smaller part of it. Thus, a two-dimensional commensurate
oscillator has a faint $SU(2)$ symmetry.
\section{The $N$-dimensional commensurate oscillator}
To describe a commensurate harmonic  oscillator in $N$ dimension,
the present notation is straightforward to adapt. Let the label
$n$ run from 1 to $N$: the Hamiltonian of the $\mathbf
m$-oscillator with ${\mathbf m} = (m_1, \ldots, m_N)$ reads
\be \label{Nosc}
H(q,p) = \frac{\omega}{2}\sum_{n=1}^N \frac{1}{m_n}(p_n^2+q_n^2)
       = \frac{\omega}{2}\sum_{n=1}^N \frac{1}{m_n}
                                      \bar{\alpha}_n\alpha_n
       = \frac{\omega}{2}\sum_{n=1}^N \frac{1}{m_n} I_n \, ;
\ee
the complex canonical variables are given by $\alpha_n =
(q_n+ip_n) / \sqrt{2}, n=1 \ldots N$, while actions $I_n$ and
angles $\varphi_n$ are defined through $ \alpha_n \sqrt{I_n} \exp
[i\varphi_n ]$. Thus there are three sets of $N$ pairs of
canonical variables to choose from, with brackets
\be \label{Nbrackets}
 \{ q_n, p_{n'} \} = \frac{1}{i} \{ \bar{\alpha}_n , \alpha_{n'} \}
                 = \{ I_n, \varphi_{n'} \} = \delta_{nn'} \, ,
                 \qquad n,n' =1 \ldots N \, .
\ee
It will be assumed that the positive integer numbers $m_n$ do not
have an overall common divisor. For the discussion to follow, two
cases will be distinguished: a commensurate oscillator is said to
be {\em canonical} if {\em no} pair of numbers $m_n$ and $m_{n'}$,
$n \neq n'$, admits a common divisor but one. This class will be
studied first. The presence of common divisors among subsets of
the frequencies $\omega_n$ gives rise to interesting additional
complications which will be considered later on.
\subsubsection*{Constants of motion and Lie algebras}
Inspired by Eq. (\ref{thirdinv}), each function
\be \label{manythirdinv}
K_{nn'} = \alpha_n^{m_n} (\bar{\alpha}_{n'})^{m_{n'}} \, ,
        \qquad n,n' =1 \ldots N \, ,
\ee
is seen to be an invariant for the commensurate $N$-oscillator, $
\{ H , K_{nn'} \} = 0$. These $N^2$ constants of motion depend on
only $N(N+1)/2$ real invariants, $N$ independent actions $I_n, n =
1 \ldots N$, and $N(N-1)/2$ relative angles
\be \label{relangles}
\chi_{nn'} = m_n \varphi_n - m_{n'} \varphi_{n'} \, ,
          \qquad 1 \leq n < n' \leq N \, .
\ee
As in the two-dimensional case, the range of the functions
$\chi_{nn'}$ must be restricted to the interval $[0,2\pi)$ because
two values $\chi_{nn'}$ and $(\chi_{nn'}+2\pi)$, respectively,
correspond to the {\em same} orbit. The angles $\chi_{nn'}$
satisfy $(N-1)(N-2)/2$ linear relations,
\be \label{angledependencies}
\chi_{nn'}+\chi_{n'n''}+\chi_{n''n}\equiv 0 \,
   \qquad n, n', n'' \, \mbox{ all different} \, .
\ee
Therefore, there are no more than $(2N-1)$ functionally
independent constants of motion, the maximum number of possibly
independent invariants. As independent invariants, one may choose,
for example, the $N$ actions $I_n$ and $N-1$ relative angles
$\chi_{n \, n+1}, n=1 \ldots  N-1$.

The $(2N-1)$- dimensional surface of constant energy $H=E$ is an
ellipsoid ${\cal E}(E)$ in phase space $\Gamma$. It contains the
$N$-dimensional torus ${\cal T}(I_1,...,I_N)$ of constant actions
$I_n$ as a sub-manifold. Lines of constant actions and angles are
the {\em orbits} of the motion, winding around a torus ${\cal T}$.
Each orbit is a one-dimensional closed loop given by
\be \label{Norbits}
\alpha_n(t) = \sqrt{I_n} \exp(-i\omega t/m_n+\varphi_n(0)) \, ,
\ee
where $m_n\varphi_n(0)-m_{n+1}\varphi_{n+1}(0)=\chi_{n,n+1}(0)$.
One revolution is completed after a time $t=2\pi M/\omega$, with
the number $M$ taking a value such that the {\em winding numbers}
$w_n=M/m_n$ of each subsystem are integer without overall common
divisor. In the canonical case, $M$ is equal to $\Pi_1^N m_n$.
Here is an example for $N=3$ which illustrates the non-canonical
case: let ${\mathbf m} = (km'_1,km'_2,m_3)$. The number $M$ would
then take the value $k m'_1 m'_2 m_3 = m_1m_2m_3/k$.

It is important to note that in a canonical (but not isotropic)
${\mathbf m}$-oscillator ({\em i.e.}, all $m_n \neq 1$), there
exist orbits with $(2^N-1)$ different periods. There are $N$
orbits corresponding to motion of a single oscillator only; there
are $N(N-1)/2$ orbits winding around two-dimensional tori with
frequencies $1/m_n$ and $1/m_{n'}, 1 \leq n < n' \leq N$, etc.

As in the two-dimensional case, the maps $R_n \alpha
=(\alpha_1,...,e^{i2\pi/m_n}\alpha_n,...,\alpha_N)$ generate a
cyclic group ${\cal C}_m = \{R_1^{r_1}\cdots R_N^{r_N} | r_n \in
{\sf Z \! \! \sf Z}\}$, the ambiguity group of the map $\psi$:
\be \label{Nambiguity}
\psi(R_1^{r_1}\cdots R_N^{r_N}\alpha) = \psi(\alpha) \,.
\ee
\subsubsection*{Reduced phase space and space of invariants}
The $(2N-1)$ phase-space functions $I_n$ and $\chi_{n \, n+1}$
form a basis of a Lie algebra commuting with the Hamiltonian $H$.
Since the functions $\chi_{nn'}$ are not continuous on phase space
${\Gamma}$, it is natural to look at appropriate periodic
functions of them. Introduce, in analogy to Eq.
(\ref{reducedvar}), the set of invariants
\be \label{Nbetas}
\beta_n = \frac{ |\alpha_n|} {\sqrt{m_n}}
           \left(\frac{\alpha_n}{|\alpha_n|}\right)^{m_n}
        = \sqrt{\frac{I_n}{m_n}}\exp[im_n\varphi_n] \, ,
          \qquad n=1 \ldots N \, .
\ee
They provide canonical coordinates on the $2N$-dimensional reduced
phase space $\Gamma_{\mathbf m}$, now with ${\mathbf m} =
(m_1,...,m_N)$,
\be \label{Ngammaredcoor}
\{\bar{\beta}_n,\beta_{n'} \} = i\delta_{nn'} \, .
\ee
As before, (\ref{Nbetas}) is a non-bijective map $\psi$: $\alpha
\to \beta(\alpha)$. It is not a projection of the phase space on a
subspace but should be thought of as a {\em ramified} cover of the
reduced space $\Gamma_{\mathbf m}$.

Not surprisingly, Eq. (\ref{condenseinv}) has a straightforward
generalization. With $\beta = ( \beta_1, \ldots,$ $ \beta_N)$, one
defines $(N^2-1)$ invariants sesquilinear in the coordinates
$\beta_n$ by
\bea \label{genuN}
J_{nn'}^s &=& \frac{1}{2} \bar{\beta} \cdot
              \left(E_{nn'} + E_{n'n}\right) \beta
           = \frac{1}{2}\left( \bar{\beta}_n \beta_{n'} +
                          \bar{\beta}_{n'}\beta_n \right) \, ,
                          \qquad 1 \leq n<n' \leq N \, ,\\
J_{nn'}^a &=& \frac{1}{2} \bar{\beta} \cdot \frac{1}{i}
               \left(E_{nn'} - E_{n'n} \right) \beta
           = \frac{1}{2i} \left( \bar{\beta}_n \beta_{n'} -
                      \bar{\beta}_{n'}\beta_n \right) \, ,
                      \qquad 1 \leq n<n' \leq N \, , \\
J_{nn}^d &=& \frac{1}{2} \bar{\beta} \cdot
             \left( E_{nn} - E_{n+1 \, n+1}\right) \beta
          = \bar{\beta}_n \beta_{n}-\bar{\beta}_{n+1} \beta_{n+1}\, ,
           \qquad n = 1 \ldots N-1 \, .
\eea
The matrices $E_{nn'}$ are of size $N \times N$ with elements
\be \label{uNgenerators}
\left(E_{nn'}\right)_{kk'} = \delta_{nk} \delta_{n'k'} \, ,
           \qquad  n,n',k,k' = 1 \ldots N \, ,
\ee
{\em i.e.}, the only nonzero elements are equal to one at position
$(n,n')$, and they generate the Lie-algebra $u(n)$ with respect to
the matrix commutator \cite{cornwell90}. This property is
inherited by the $N^2$ phase-space functions
\be \label{Npoissonbrackets}
J_{nn'} = \bar \beta \cdot E_{nn'} \beta
        \equiv \bar \beta_n \beta_{n'} \, ;
\ee
their Poisson brackets,
\be \label{complexuN}
\{J_{nn'} , J_{kk'}\}
  = i \left( \delta_{nk'} J_{kn'} - \delta_{n'k} J_{nk'} \right) \, ,
\ee
also realize the algebraic relations of $u(N)$.

It is possible to find $(N^2-1)$ linear combinations of the
matrices $E_{nn'}$ which are traceless and hermitean---hence they
provide a basis of the algebra $su(N)$. In fact, these
combinations have been introduced already in Eq. (\ref{genuN})
when defining  $J_{nn'}^s, J_{nn'}^a$, and $J_{nn}^d$. Therefore,
these functions form a basis of the algebra $su(N)$ with respect
to the Poisson bracket. When supplemented by (a multiple of) the
Hamiltonian
\be \label{Nhamilton}
J_0 = \frac{1}{2} \bar \beta \cdot {\bf 1}_N \beta
    = \frac{1}{2} \sum_{n=1}^N \bar \beta_n \beta_n
    \equiv \frac{2}{\omega} H \, ,
\ee
where ${\bf 1}_N$ is the $N$-dimensional unit matrix, the algebra
$u(n)$ can be recovered.
\subsubsection*{Vectorfields}
There is a first group of transformations which acts in the space
of invariants $\Upsilon$. As before, it is the set of finite
transformations on the space generated by the real invariants
(\ref{genuN}). In other words, it arises from integrating the
adjoint representation of the algebra formed by the invariants. As
this group will play no role in the following, its discussion is
suppressed.

Next, the invariants $J_{nn'}^s$ and $J_{nn'}^a$, generate
canonical linear maps in the reduced space $\Gamma_{\mathbf m}$,
\begin{eqnarray} \label{Nflows}
\frac{d\beta_k}{d\tau}
  & = & \{\beta_k , J_{nn'}^s \}
    = \frac{i}{2}(\delta_{kn}\beta_{n'} + \delta_{kn'}\beta_{n})
    \equiv (J_{nn'}^s\beta)_k \, ,\\
\frac{d\beta_k}{d\tau}
  & = & \{\beta_k,J_{nn'}^a \}
    = \frac{1}{2}(\delta_{kn}\beta_{n'} - \delta_{kn'}\beta_n)
    \equiv (J_{nn'}^a\beta)_k \, ,
\end{eqnarray}
and similar ones follow when taking $J_{nn}^d$ as generator. These
linear equations can be integrated in the space $\Gamma_{\mathbf
m}$ for arbitrary initial values $\beta(0)=\beta_0 \in
\Gamma_{\mathbf m}$,
\be \label{Nsolutions}
\beta(\tau) = \exp(\tau J_{nn'}^\varepsilon) \beta_0 \, ,
              \qquad \varepsilon =a,s \, .
\ee
The solutions are unitary maps of $\Gamma_{\mathbf m}$ to itself.
In analogy to the two-dimensional case, they will be denoted by
\be \label{geomnotation}
\beta(\tau)=\mbox{Exp}[\tau J_{nn'}^\varepsilon] (\beta) \, ,
            \qquad \varepsilon =a,s \, .
\ee
and similar for finite transformations generated by the invariants
$J_{nn}^d$. Due to the linearity of the equations, no ambiguities
arise upon integration. Therefore, the set of transformations in
the reduced space $\Gamma_{\mathbf m}$ is isomorphic to the group
$SU(N)$. In this restricted sense, the ${\mathbf m}$-oscillator
has the special unitary group in $N$ dimensions as a symmetry
group. This group of symmetry transformations discussed is {\em
not} defined in the phase space $\Gamma$ of the ${\mathbf
m}$-oscillator.

Finally, a genuine `pullback' of $SU(N)$ in phase space ${\Gamma}$
does not exist, for the same reasons as in the case $N=2$. In
fact, it is sufficient to consider a pair of oscillators with
frequencies $\omega/m_n$ and $\omega / m_{n'}$, say, in order to
see that there are obstructions which prevent the existence of a
{\em global} symmetry group in phase space $\Gamma$. This pair of
degrees of freedom is equivalent to a two dimensional
$(m_n,m_{n'})$-oscillator, and no set of transformations acting on
it can be found which would be isomorphic to $SU(2)$. If, however,
the $N$-dimensional oscillator would have the full symmetry
$SU(N)$, a subgroup $SU(2)$ should be associated with this pair of
oscillators. Consequently, the group $SU(N)$ cannot be identified
as symmetry group of the canonical $N$-dimensional commensurate
oscillator. In analogy to the two-dimensional commensurate
oscillator it is seen to have a {\em faint} $SU(N)$ symmetry only.
\subsubsection*{The ${\mathbf m}$-oscillator with common divisors}
The canonical ${\mathbf m}$-oscillator has been shown to be
invariant under transformations isomorphic to the group $SU(N)$ in
the reduced space ${\Gamma}_{\mathbf m}$. For canonical and
isotropic $N$-dimensional oscillators, subsystems of dimension
$N'<N$ are invariant only with respect to a subalgebra ${\cal
A}_{{\mathbf m}'}$ of the algebra ${\cal A}_{\mathbf m} = su(N)$.
If the oscillator is neither isotropic nor canonical, other
possibilities arise.

A non-canonical oscillator is characterized by frequencies
${\mathbf m} = (m_1,...,m_N)$ with at least one pair $(m_k,m_l)$
having a common integer divisor different from one.  Let $N'<N$
frequencies have a common divisor. Then, for the ${\mathbf
m}'$-oscillator corresponding to these frequencies, constants of
motion do exist which form an algebra ${\mathcal A}_{{\mathbf m}'}
= su(N')$. This algebra, however, is {\em not} a subalgebra of
${\mathcal A}_{\mathbf m}$ as follows immediately from considering
the ${\mathbf m}'$-oscillator as an $N'$-dimensional commensurate
oscillator in its own right. Suppose that, after removing the
common divisor, the resulting oscillator, characterized by
${\mathbf m}' = (m'_1, \ldots , m'_{N'})$, is a canonical one.
Then one can construct a group of symmetry transformations
$SU(N')$ in the reduced phase space $\Gamma_{{\mathbf m}'}$, and
$\Gamma_{{\mathbf m}'}$ is {\em not} a subspace of
$\Gamma_{\mathbf m}$. The Poisson brackets of the generators of
$SU(N')$ acting in $\Gamma_{{\mathbf m}'}$ and those of $SU(N)$
acting in $\Gamma_{\mathbf m}$ will not be linear combinations of
the initial ones. Hence, the combination of these two algebras
will not close under the Lie product---the resulting algebra will
be {\em infinite}. This property will be important for quantum
mechanical commensurate oscillators since it entails additional
degeneracies of energy which otherwise appear to be accidental.

Turn these results around: there is no finite algebra to account
for all the symmetries of a non-canonical ${\mathbf
m}$-oscillator. Obviously, this situation can arise only if $N
\geq 3$ (if $N=2$ any common divisor can be factored out
immediately). In fact, if $m_{n'}^{'} = 1$, $n' = 1 \ldots N'$,
the subsystem is even an isotropic oscillator, and it has a group
$SU(N')$ of symmetry transformations on phase space $\Gamma$.

It is helpful to illustrate this discussion by an exhaustive list
of `classes' for small values of $N$.
\begin{itemize}
\item[$\diamondsuit$]
$N=2$: A commensurate oscillator is either isotropic or canonical
(a common divisor of the frequencies $m_1 \neq m_2$ can be
factored out).
\item[$\diamondsuit$]
$N=3$: Five classes of commensurate oscillators can be identified.
An ${\mathbf m}$-oscillator is either isotropic or canonical, or
it belongs to one of the three following classes:
\begin{itemize}
\item[1.] a single pair of two frequencies have a common divisor,
          ${\mathbf m}=(jm'_1,jm'_2,m_3)$, say;
\item[2.] two pairs have common but different divisors,
          ${\mathbf m}=(jm'_1,jkm'_2,km'_3)$, say;
\item[3.] all three pairs have common but different divisors,
          ${\mathbf m}=(jkm'_1,klm'_2,ljm'_3)$, say.
\end{itemize}
\end{itemize}
For $N>3$ the number of different classes increases rapidly with
$N$.

Consider an example of Type 1 for $N=3$ in detail. The three
coordinates $\beta_n$ of the space ${\Gamma}_{\mathbf m}$ allow
one to define eight constants of motion $J$. In addition,
introduce coordinates of the reduced phase space
${\Gamma}_{{\mathbf m}'}$,
\be \label{primedreducedbeta}
 \beta'_n = \frac{|\alpha_n|}{\sqrt{m_n'}}
            \left(\frac{\alpha_n}{|\alpha_n|} \right)^{m_n'} \, ,
            \qquad n=1,2 \, .
\ee
The four functions
\be \label{noncanonconst}
J'_{nn'} = \bar{\beta}'_n \beta'_{n'} \, ,
          \qquad n,n'=1,2 \, ,
\ee
are a different set of constants of motion because the Hamiltonian
of the subsystem $(1,2)$ has an overall factor $1/k$. The
constants $J'$ are the basis of a Lie algebra ${\mathcal
A}_{{\mathbf m}'}$ isomorphic to $su(2)$ (setting aside the fourth
commuting invariant), as the subsystem is an ${\mathbf
m}'$-oscillator with $N=2$. The resulting algebra ${\mathcal
A}_{{\mathbf m}'}$ gives rise to another faint $SU(2)$ symmetry.
It is, however, {\em neither} a subalgebra of the faint $SU(N)$
symmetry (as it it implemented on a different reduced phase space
$\Gamma_{{\mathbf m}'}$) {\em nor} do the generators of ${\mathcal
A}_{\mathbf m}$ and ${\mathcal A}_{{\mathbf m}'}$ commute.
Consequently, the union of both algebras gives rise to an infinite
algebra. Finally, if $m_1^{'} = m_2^{'} = 1$, three of the
functions $J'$ would generate the group $SU(2)$ on the original
phase space $\Gamma$. In other words, the faint $SU(N)$ symmetry
of an $\bf m$-oscillator with common divisors is compatible with
the existence of smaller groups acting globally in phase space
$\Gamma$.
\section{Summary and Outlook}
This paper deals with the problem which symmetry group to
associate with an $N$-dimensional {\em commensurate} harmonic
oscillator. Structural similarities seem to indicate that the
introduction of rational frequency ratios $m_n/m_{n'}$ would not
affect the existence of the group $SU(N)$ as  a group of symmetry
transformations. This suggestion was based on the following
observations.  Arbitrary rational frequency ratios $m_n/m_{n'}$,
are still compatible with the existence of $(2N-1)$ globally
defined invariants. In both cases, the invariants confine
trajectories to a one-dimensional manifold in phase space, the
orbit. Furthermore, the invariants form an algebra $su(N)$ with
respect to the Poisson bracket. There is, however, a subtle
difference between an isotropic and a commensurate oscillator:
isotropy forces all orbits to have the {\em same} period whereas
commensurate frequencies allow for orbits with {\em different}
periods. Consequently, these system are distinguishable from an
experimental point of view.

It has been shown that the algebra $su(N)$ of the commensurate
oscillator cannot be extended globally to a representation of the
group $SU(N)$ in phase space. Strictly speaking, it is thus not
possible to attribute this group as a symmetry group to the
commensurate harmonic oscillator. The group $SU(N)$ is associated
with commensurate oscillators in a restricted sense only: to do
so, the action of the invariants must be considered in a {\em
reduced} phase space the points of which are no longer in a
one-to-one correspondence with the states of the system. The
commensurate oscillator is said to have a {\em faint} $SU(N)$ {\em
symmetry}. Furthermore, if the rationally related frequencies have
common divisors, additional sets of symmetry transformations can
be found. They are not subgroups of the faint group $SU(N)$, which
acts in reduced phase, but they act in different reduced phase
spaces.

To conclude, it has been shown that the symmetries of commensurate
harmonic oscillators come in a surprisingly rich variety and
depends in a subtle way on the frequency ratios. Classical and
quantum mechanical oscillators are closely related. Therefore, it
will be promising to study the impact of faint symmetries on the
Hilbert-space structure of quantum mechanical commensurate
oscillators \cite{amiet+02}. In particular, a systematic
group-theoretical account of their degenerate energy levels is
expected to benefit from the concept of faint symmetry.
\subsubsection*{Acknowledgements}
St. Weigert is grateful for support by the Swiss National Science
Foundation.


\begin{thebibliography}{10}
%
\bibitem{perelomov90}
          A.M. Perelomov:
          {\sl Integrable Systems of Classical Mechanics
          and Lie Algebras I.}
          Basel, Boston, Berlin: Birkh\"auser 1990
\bibitem{arnold}
         V. Arnold:
         {\sl M\'{e}thodes Math\'{e}matiques de la M\'{e}canique
         Classique.} Moscou: Mir 1976
\bibitem{amiet+02}
         J.-P. Amiet and St. Weigert:
         {\em Commensurate Harmonic Oscillators: Quantum Symmetries}
         (in preparation)
\bibitem{jauch+40}
         J.M. Jauch, and E.L. Hill:
         Phys.\ Rev.\ {\bf  57} (1940) 641
\bibitem{ae}
         W. Pauli:
         ZS.\ f.\ Phys. {\bf 36} (1926) 336
\bibitem{hulthen33}
         See the footnote in L. Hulth\'{e}n:
         ZS.\ f.\ Phys.\ {\bf 86} (1933) 21
\bibitem{mcintosh59}
         H.V. McIntosh:
         Am.\ J.\ Phys.\ {\bf 27 } (1959) 620
\bibitem{d} V.A. Dulock, and H.V. McIntosh:
         Am.\ J.\ Phys.\ {\bf 33 } (1965) 109
\bibitem{f}
         G. Maiella, and B. Vitale:
         Nuov.\ Cim.\ {\bf 47} A (1967) 330
\bibitem{g}
         G. Maiella:
         Nuov.\ Cim. {\bf 52 } A (1967) 1004
\bibitem{ai}
         J. Moser:
         Commun.\ Pure Appl.\ Math.\ {\bf 23} (1970) 609
\bibitem{o}
         H. Bacry, H. Ruegg, and J.-M. Souriau:
         Commun.\ math.\ Phys.\ {\bf 3} (1966) 323
\bibitem{l}
         V.A. Dulock, and H.V. McIntosh:
         Pacif.\ J.\ Math.\ {\bf 19}(1966) 39
\bibitem{p}
         D.M. Fradkin:
         Prog.\ Theor.\ Phys.\ {\bf 37} (1967) 798
\bibitem{r}
         N. Mukunda:
         Phys.\ Rev.\ {\bf 155} (1967) 1383
\bibitem{s}
         N. Mukunda:
         J.\ Math.\ Phys.\ {\bf 8} (1967) 1069
\bibitem{aj}
         L.P. Eisenhart:
         {\sl Continuous Groups of Transformation.}
         New York: Dover Publications 1963
\bibitem{u}
         P. Stehle, and M.Y Han:
         Phys.\ Rev.\ {\bf 159} (1967) 1076
\bibitem{v}
         M.Y. Han, and P. Stehle:
         Nuov.\ Cim.\ {\bf 48} (1967) 180
\bibitem{moshinsky+78}
         M. Moshinsky and T. H. Seligman:
         Ann.\ Phys.\ (N.Y.) {\bf 114}, 243-272 (1978)
\bibitem{cornwell90}
         J.F. Cornwell:
         {\sl Group Theory in Physics.} Vol. II.
         London: Academic Press $^4$1990.
%
\end{thebibliography}
\end{document}